\documentclass[fleqn,10pt]{wlscirep}

\usepackage{fourier}
\usepackage[utf8]{inputenc}
\usepackage{graphicx}
\usepackage{amsmath}
\usepackage{caption}
\usepackage{verbatim}
\usepackage{cuted}
\usepackage{subfig}

\title{Design of Oscillatory Neural Networks by Machine Learning}

\author[1]{Tam\'{a}s Rudner}
\author[2]{Wolfgang Porod}
\author[1]{Gy\"{o}rgy Csaba}
\affil[1]{P\'{a}zm\'{a}ny P\'{e}ter Catholic University, Faculty of Information Technology and Bionics, Budapest, Hungary}
\affil[2]{Center for Nano Science and Technology University of Notre Dame (ND\textit{nano}), Notre Dame, IN, USA}

\affil[*]{gcsaba@gmail.com}

\begin{abstract}

We demonstrate the utility of machine learning algorithms for the design of Oscillatory Neural Networks (ONNs). After constructing a circuit model of the oscillators in a machine-learning-enabled simulator and performing Backpropagation through time (BPTT) for determining the coupling resistances between the ring oscillators, we show the design of associative memories and multi-layered ONN classifiers. The machine-learning-designed ONNs show superior performance compared to other design methods (such as Hebbian learning) and they also enable significant simplifications in the circuit topology. We demonstrate the design of multi-layered ONNs that show superior performance compared to single-layer ones. We argue Machine learning can unlock the true computing potential of ONNs hardware. 

\end{abstract}
\begin{document}

\flushbottom
\maketitle
\thispagestyle{empty}

\section*{Introduction}
\label{Intro}

Throughout the history of electronics, analog computing devices were often looked at as the future of computing - only to be turning up to be runner-ups behind their digital counterparts \cite{ref:bournez}. Despite this history, research on analog computing is again on the rise, for multiple reasons. With the end of Dennard scaling \cite{ref:moore}, transistor resources are less plentiful and analog devices fit much better to sensory preprocessing functions then digital ones. New sensory processing pipelines require less and less digital number crunching and more neuromorphic, edge-AI functions, where analog computing devices may have an edge. For this reason, research is flourishing in analog hardware accelerators \cite{ref:superimportant} (aka neuromorphic hardware devices) that have the potential to boost the energy efficiency of AI processing pipelines by several orders of magnitude.

Among the many flavors of analog computing, Oscillatory Neural Networks (ONNs) received special attention \cite{ref:csaba_perspectives}. This is due to the facts that (1) ONNs are realizable by very simple circuits - either by emerging devices or CMOS devices (2) phases and frequencies enable a rich and robust \cite{{ref:csaba_perspectives}} representation of information (3) biological systems seem to use oscillators to process information \cite{ref:furber_temple} - likely for a reason.

Despite the significant current research efforts and the large literature, most ONNs seem to rely on some version of a Hebbian rule to define attractor states \cite{ref:aida_hebbian}, so that the network converges to a stationary phase pattern, which is a the result of the computation. The Hebbian rule is used to calculate the value of physical couplings that define the circuit function. The reliance on the Hebbian rule turns ONNs to a sub-class of classical Hopfield networks - which are not very powerful by today's standards. While there are a few ONN implementations not relying on basic Hebbian rules (notably \cite{ref:suppes}) it is likely that current ONNs does not fully exploit the potential of the hardware - due to the lack of a more powerful method to design the interconnections.

In this paper we show that a state of art machine learning method, when applied to a SPICE-like model of the circuit, significantly enhances the computational power of  ONNs. Our studied system is an ONN made of resistively coupled ring-oscillators \cite{{ref:montecarlo}}, \cite{ref:kim} - its circuit topology is described in \ref{sec:ringosc}. Next, in Section \ref{sec:method} we develop the differential equations describing the circuit and show how a machine learning algorithm can be applied to design the circuit parameters. In Section \ref{sec:mnist}, we exemplify the machine-learning framework for the design of an auto-associative memory and compare it to a standard Hebbian-rule based device. Section \ref{sec:classifier} furthers this concept by the design of a multi-layered network, which is a two-layer classifier and achieves superb performance compared to a single-layer device. 

Overall our work presents a design methodology that unlocks the true potential of oscillatory neural networks, overcoming the limitations of imposed by simple learning rules. Additionally, the presented method allows for designing physically realizable structures: our networks rely on nearest-neighbor interactions, which is amenable to scaling, chip-scale realizations and uses significantly fewer neurons than fully connected networks.

%This latter network bears no direct analogy to Hebbian autoassociatve memories and shows the potential of the method to design multileyered, complex ONNs. 

\section{Resistively coupled ring oscillators for phase-based neuromorphic computation}
\label{sec:ringosc}

It is well established that the synchronization patterns of coupled oscillators may be used for computation. The idea of using phase for Boolean computation goes back to the early days of computer science \cite{{ref:vonneumann}} and is being rediscovered this days \cite{ref:rowch}.  For neuromorphic computing the original scheme of Izhikevich \cite{{ref:izhikevich}}, \cite{ref:izhikevich2} was studied using various oscillator types  and coupling schemes. A number of computing models were explored, ranging from basic convolvers \cite{ref:upside} and pattern generators \cite{ref:suman_cpg} to hardware for handling NP-hard problems \cite{ref:kim}, \cite{ref:graphcoloring_earliest}, \cite{ref:suman_coloring}.

\begin{figure}[!ht]
    \centering
   \includegraphics[width=0.8\textwidth]{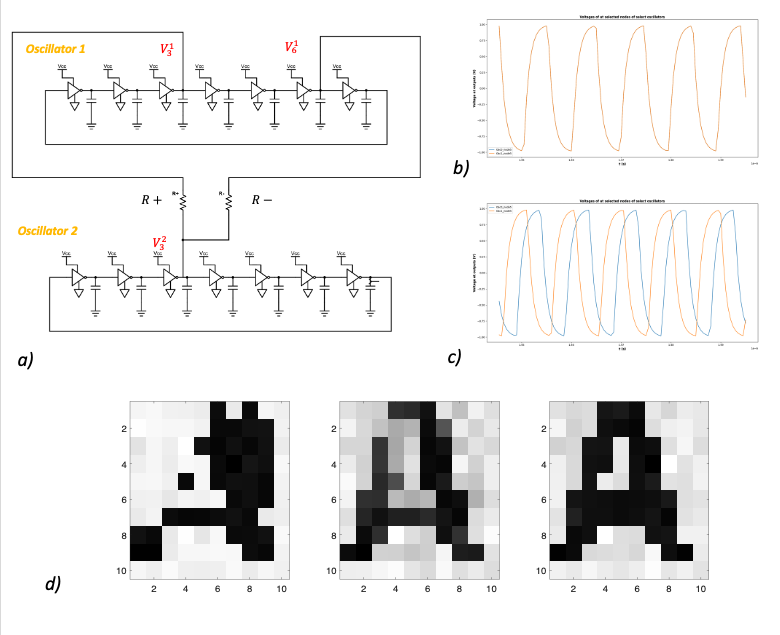}
    
    \caption{Phase-based computing by two ring oscillators. $a)$ a network topology is shown b) if $R^+$ dominates in the coupling, the oscillators run in phase, while if  $R^+$ dominates ($b)-c)$) then anti-phase coupling is realized. $d)$ If phases corresponds to pixels of a greyscale image, the phase dynamics may be used to converge to preprogrammed patterns \cite{ref:montecarlo} }%
    \label{fig:oscoupling}%
\end{figure}

To give a simple example of how ring oscillators compute in phase space, Fig. \ref{fig:oscoupling} shows a simple two-oscillator system. Nodes which are interconnected by a resistor will synchronize in-phase. If identical nodes (say $V_3$) are interconnected, the oscillators will run in phase.  However, in a 7-inverter ring oscillators, each node is phase-shifted by an angle of $2\pi/7$ with respect to their neighbor. So if say say $V_3$ of one oscillator is connected to say $V_6$ of the oscillators, the oscillators will pull toward an anti-phase configuration. The waveforms of these two cases are illustrated in Fig. \ref{fig:oscoupling}$b)$-$c)$.

A larger network of oscillators with in-phase or out-of phase pulling resistors will converge toward an oscillatory ground state configuration, which in fact maps to the solution of the Ising problem \cite{ref:kim}. Simply put, the phase of each oscillator will converge toward a value that optimally agrees to most of the constraints imposed on the oscillator, and the dynamics of the coupled oscillator network will approximate the solution of a computationally hard optimization problem. For an Ising problem, the oscillator-oscillator couplings are part of the problem description, there is no need to calculate them.

While the Ising problem is important and shows the computational power of ONNs, an Ising solver alone is not very useful for solving most real-life, neuromorphic computing tasks. A neurmorphic computing primitive (such as a classification task) does not straightforwardly map to an Ising problem. So if the oscillator network is to be used as a neuromorphic hardware, then the oscillator weights must be designed or learned to perform certain computational functions. 

Most ONNs are used as auto-associative memories, making them applicable for simple pattern recognition / classification tasks. The weights are designed based on the Hebbian learning rule \cite{ref:montecarlo}, \cite{ref:aida_hebbian} - and this is one of the cases when the Ising model easily maps to a neuromorphic computing model. In fact, the connection between Ising and Hopfield associative models \cite{ref:hopfield}, \cite{ref:hopfield_comb}, \cite{ref:porod_hopfield} were done by Hopfield early on \cite{ref:hopfield_tank}. ONNs simply use oscillator phases as the state variable of Hopfield neurons.

The Hebbian rule (and even its improved variants \cite{ref:tolmachev}, \cite{ref:righetti}) has severe limitations: the rule assumes all-to-all oscillator (neural) connections and they do not support learning on a set of training examples. Also, simple Hopfield models are not very powerful neural networks by today's standards. This is why our goal in this paper is to go beyond these limitations and apply state of art learning techniques to train ONN weights. This allows us to overcome the limitations of associative (Hopfield) type models and design ONN versions of many other neural network models.

\section{Machine learning framework for circuit dynamics}
\label{sec:method}

Out methodology is to apply Backpropagation through time (BPTT) \cite{ref:bptt} to an in-silico model of the oscillators. We construct a circuit model of the coupled oscillator system, the resulting ODEs are solved and the value of the loss function is calculated at the end of the procedure. By backpropagating the error, we can optimize the circuit parameters in such a way that the ONN solves the computational task that is defined by the loss function. Once the circuit parameters are determined via this algorithm they can be 'hard wired' into a circuit (ONN hardware) for an effective hardware accelerator tool.

\subsection{Computational model of resistively coupled ring oscillators}

For the sake of concreteness we assume that our circuit consists of $n$ oscillators and each oscillator is composed of 7 inverters. The circuit has $k$ input nodes. We construct a simple Ordinary Differential Equation (ODE)-based circuit model based on the equations derived in \cite{ref:lai}. Each inverter is described on a behavioral level by a $\tanh()$ nonlinearity connected to an $RC$ delay element. This way a seven-inverter ring oscillators is modeled by seven first-order nonlinear ODEs.

The mathematical formulation consists three parts: internal dynamics of the oscillators (due to the inverters), dynamics due to external signals (inputs) and the coupling's dynamics. We can arrive at the following ODE for the collection of voltages at all the nodes, which describes all parts if we write a differential equation for every nodes in the system using Kirchhoff's current law and assuming only resistors as couplings:

\begin{equation*}
        \frac{dV}{dt} = \frac{1}{RC}\Big(f\big(\textbf{P}_{\pi}V\big) - V\Big) + \frac{1}{C}\textbf{B}'u + \frac{1}{R_cC}\textbf{C}'V, 
\end{equation*}

where 
\begin{equation*}
    f(x) = -\tanh(ax), 
\end{equation*}

is the simplified characteristic of an inverter with some $a \in \mathbb{R}$. Furthermore, $\pi$ is a permutation, such that
\begin{equation*}
    \pi = \begin{pmatrix}
            1 & 2 & 3 & 4 & 5 & 6 & 7\\
            7 & 1 & 2 & 3 & 4 & 5 & 6
          \end{pmatrix}
\end{equation*}

and $\textbf{P} \in \mathbb{R}^{(7n)\times(7n)}$ is a block matrix for which every $7x7$ matrix-block in the main diagonal there is a permutation matrix corresponding to $\pi$. Basically this orders the voltage nodes in the ring oscillator to calculate the voltage differences arising between the two endpoints of the resistors placed in-between two inverters. $\textbf{B}' \in \mathbb{R}^{(7n)\times k}$ is the connector matrix for the inputs. The inputs are collected in $u \in \mathbb{R}^k$. $\textbf{C}' \in \mathbb{R}^{(7n)\times(7n)}$ is the modified couplings matrix which is to be constructed from the real, humanly readable couplings matrix $\textbf{C} \in \mathbb{R}^{n\times n}$. The parameters $R$, $C \in \mathbb{R}^+$ are fixed for the oscillators, meanwhile the $R_c \in \mathbb{R}^+$ coupling parameters are one of the two real, to-be-learnt parameters of the system which governs the whole coupling dynamics. The other ones are the parameters gathered in $\textbf{B}'$, which is directly relates to the amplitude of the input signal (typically a sinusoidal current generator).

The $\textbf{C}_{i,j}$ is related to the couplings between oscillators $i$ and $j$ and the matrix is built the following way:
\begin{itemize}
    \item All main diagonal entries are $0$, as no oscillator is coupled to itself
    \item All entries in the upper triangle of the matrix are corresponding to the positive (in-phase-pulling) couplings
    \item All entries in the lower triangle of the matrix are corresponding to the negative (anti-phase-pulling) couplings
\end{itemize}

The construction of $\textbf{C}'$ can be done easily from $C$ algorithmically. As every positive couplings is between 3-3 nodes of the oscillators and every negative connection is between 3-6 nodes of oscillators, the $\textbf{C}'$ matrix is quite sparse. Similarly, because inputs are only fed into the 3rd node of every oscillator, the $\textbf{B}'$ matrix is sparse aswell. 

\begin{figure}[!ht]
    \centering
    \includegraphics[width=0.8\textwidth]{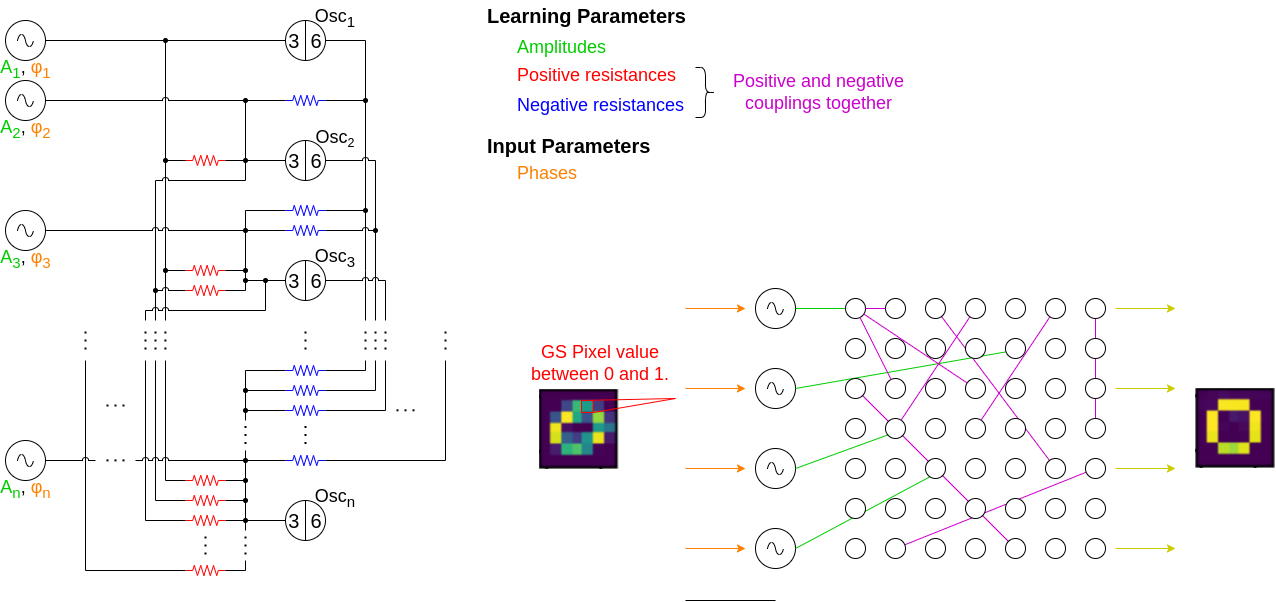}
    \caption{The circuit diagram of the entire computational layer. Input signal generators provide the sinusoidal signals with a phase that corresponds to an input pattern, such as pixels of an image. These generators are connected to the computing oscillators, whose phase pattern provides the solution of the problem. The 3-6 marks on the oscillators indicate the 3rd and 6th nodes in the ring oscillators' circuit. The green, red and blue colored circuit elements' values are learned during the learning process and the phases, indicated with orange are the inputs. On the schematic figure, the purple connections indicate both the positive (red) and negative (purple) couplings. The grayscale pixel value is read from the image, converted to phase information, then the sinusoidal current generators are connected to the oscillators one-by-one. The yellow arrows shows that the output is read from the oscillators and an image is formed.}%
    \label{fig:circuit}%
\end{figure}

The ODEs are constructed for the circuit of Figure \ref{fig:circuit} in case of a fully connected ONN. The oscillators are driven by sinusiodal current generators - the phase of these signals carry the input. They define the initial states of the oscillators that is later changed by the couplings between the oscillators.

Each oscillator is connected by two resistors, the value of which has to be learned. The value of the coupling resistors is directly related to the coupling parameters, which is stored in the $\textbf{C}$ coupling matrix. In the equations above $R_c$ is a predefined, constant value which is the baseline resistance between two coupled nodes, usually around $10 k\Omega$. The system learns the values in $\textbf{C}$ from which the $\textbf{C}'$ modified coupling matrix is built and then the learn values in $\textbf{C}$ are going to be scalers to this $R_c$ parameter, so the real, physical value of a coupling resistance between node $i$ and $j$ is given by $\frac{R_c}{\textbf{C}_{i,j}}$. In other words, the value of the learn parameters in $\textbf{C}_{i.j}$ is inversely proportional to the resistance value between oscillators $i$ and $j$.

Similarly, the values in $\textbf{B}'$ are related to the input current generator's amplitude, but they are directly proportional to the real amplitude of input generators.

In the examples of the later sections, the grayscale pixel colors will typically correspond to the input phases of the current generators - a pixel intensity from 0 to 1 is mapped to phases $\phi \in [0, \pi]$. Similarly, the output pattern is the stable, stationary phase pattern of the oscillators. 

\subsection{Backpropagation for ONN circuit design}

Backpropagation is the de facto standard algorithm used for the training of neural networks \cite{ref:deep}. After each run of the neural network, the gradient of a properly defined loss function is computed with respect to the leaenable parameters of the system, in an efficient manner. After the calculation of the gradient, a gradient descent method is applied to the learning, in order to minimize the loss function. 

BPTT (Backpropagation Through Time) is the backpropagation applied to a dynamic system (i.e. an ODE-based description). The time-discretized model of the ODEs is unfolded in time, so that one neural layer corresponds to a temporal snapshot of the system dynamics. The loss function is typically defined on the end state of the ODEs - from this the optimal value of the ODE parameters (and/or the ODE initial conditions) can be determined. We use BPTT to determine the optimal value of circuit parameters in the ring oscillator network.

We have written our simulation code in Pytorch \cite{ref:pytorch} - the autograd feature of Pytorch makes the implementation of backpropagation and BPTT straightforward. We also used the \textit{torchdiffeq} \cite{ref:torchdiffeq} package for implementing backward-differentiable ODE solvers. This external, third-party library is built upon PyTorch and provides various differentiable ODE solvers implemented for PyTorch. A particularly useful feature of \textit{torchdiffeq} is that it can apply the adjoint method for the backward step \cite{ref:adjoint}, and calculate the gradients with a constant memory cost.

\begin{figure}[!ht]
    \centering
    \includegraphics[width=0.9\textwidth]{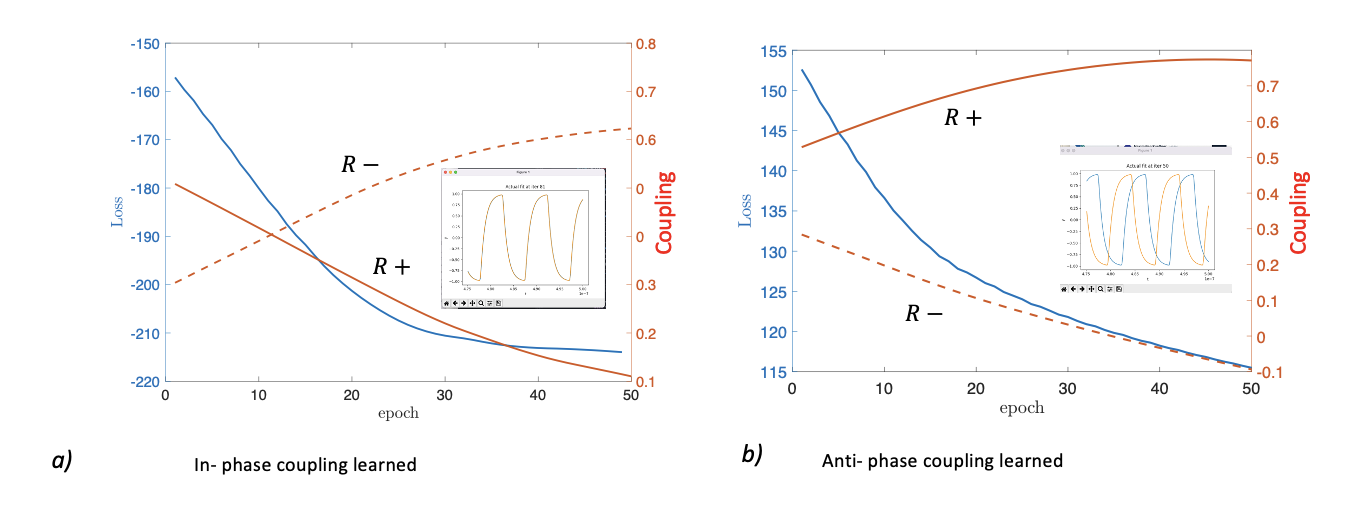}
    \caption{On a), we can see the simulation's result for the positively coupled oscillators, meanwhile on b), there is the same for the negatively coupled 2-oscillator system. The loss changed in both cases from high value to low value. Also, the orange curves are indicating the learn parameters values contained in $\textbf{C}$ and not the real values of the resistors. Note that in b) the parameter value corresponding to "R-" is going below 0, which would mean a negative resistance because of the connection of the parameters in $\textbf{C}$ to the physical parameters, but this is only the mathematical solution, for a given simulation the parameters were clamped to be non-negative and if they hit zero, the connection  removed.}
    \label{fig:osc2learning}
\end{figure}

Figure \ref{fig:osc2learning} exemplifies the learning procedure for the two-oscillator system of Fig. \ref{fig:oscoupling}. We defined the loss function of the system as the dot product of the oscillator waveforms - which should be maximized (minimized) for in-phase (anti-phase) coupling. The machine learning algorithm adjusts the value of the $C_{i,j}$ parameters (and the coupling resistors) until this desired phase configuration is reached.

This method can be straightforwardly generalized to achieve convergence toward more complex  patterns. If the loss function aims to maximize the dot product of waveforms between like-colored pixels and minimize them between different-colored ones, then the phase pattern can converge toward any prescribed image. If the phase pattern made to converge toward different patterns for different inputs, then the ONN will act as an associative memory.

Since the Machine Learning (ML) technique is designing a physical circuit, safeguards was taken not to arrive to unrealizable circuit parameters such as negative resistances or exceedingly strong couplings that would quench oscillations.

\section{ONN-based pattern classification on the MNIST dataset}
\label{sec:mnist} 

We have chosen the standard MNIST database for testing the associative capabilities of our system. Since the BPTT algorithm is computationally demanding, we made a few simplifications. We downsampled the initially 28x28 pixel-sized picture from MNIST to have either 14x14 or 7x7 size using average pooling. This allowed us to have a reduced dimension for the input images but also keeping the necessary information because of the average pooling. Also, 14x14 MNIST images are still recognisable as a human, so it allowed us to easily recognise if some patterns are easier for the algorithm to distinguish from the others. 

\subsection{Baseline: ONN-based associative memory with Hebbian learning}

The simplest, well studied ONN based associative memory can be designed by the Hebbian rule.  If we want the phase pattern to converge toward $\xi$ or $\eta$ for inputs resembling to $\xi$ or $\eta$, then the weights that realize this associative memory are:

\begin{equation}
    C^{cpl}_{ij} = \dfrac{1}{2} \big(\xi_i\xi_j + \eta_i\eta_j\big), 
\end{equation}

where $\xi_i$, $\xi_j$ is the $i$-th, $j$-th element of the pattern $\xi$, and $\eta_i$, $\eta_j$ is the $i$-th, $j$-th element of the pattern $\eta$, respectively.

The rule assumes all-to-all couplings, making a larger-scale network hard to physically realize.

Hebbian learning is not an iterative learning process, the weights are determined in a single-shot formula. To improve the results we applied machine learning to optimize the value of base coupling resistances, $R_c$ and the parameters in $\textbf{B}'$, or in other worlds, the amplitudes of input current generators.

The inner $RC$ time constant of the Ring-oscillators was $2.0\cdot 10^{-10}$ sec, which translates into a 500 MHz oscillation frequency (time period $T=2$ ns). The total simulation time for the network is 500 ns. The phase pattern is calculated from the last 300 ns window, so convergences is achieved after less then 100 oscillation cycles or 200 ns.

\subsection{ONN-based associative memories designed with machine learning}

The same functionality that is realized by Hebbian learning can be achieved by BPTT method. The loss function we designed was:

\begin{equation}
    L = \dfrac{1}{n} \sum_{k = 0}^{n} (O_n - T_n)^2,
\end{equation}

where $O_k$ is the pattern calculated from the output of the oscillators for the $k$-th input in the batch and $T_k$ is the ground truth for the same, which were ideal patterns of '0' and '1'. In the above formula $n$ is the size of the batch used for learning.

\begin{figure}[!ht]
    \centering
    \subfloat[Proposed fully-connected circuit design by ML]{{\includegraphics[width=8cm]{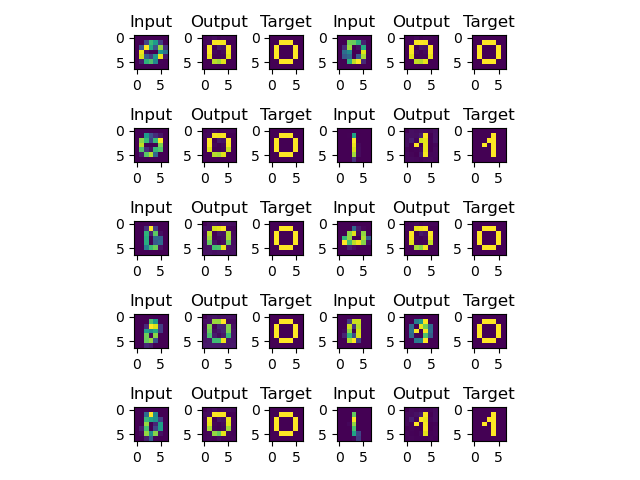} }}%
    \qquad
    \subfloat[Circuit with existing Hebbian-learning based couplings]{{\includegraphics[width=8cm]{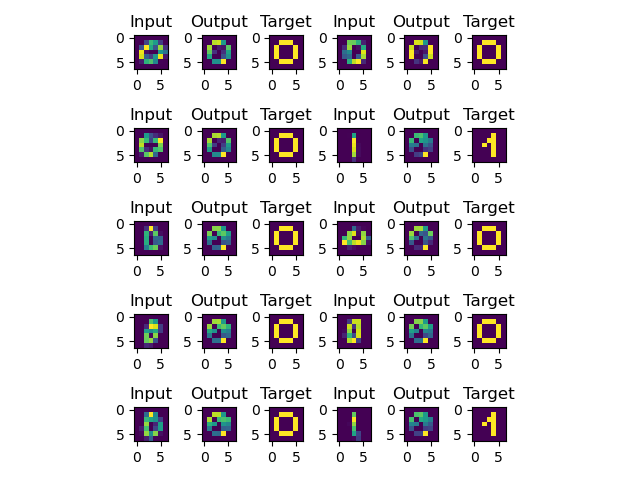} }}%
    \caption{On a) we can see the results of the fully connected network trained by ML, while on b) we can see the results of the system trained by the Hebbian-learning scheme. It is evident, that the former performs better qualitatively, and also, we calculated the mean-squared errors for the set, and for the proposed network, it was 0.020, meanwhile for the Hebbian it was 0.068, so our solution was more than 3 times better.}%
    \label{fig:test_compare}%
\end{figure}

Figure \ref{fig:test_compare} compares results from tthe Hebbian and BPTT based designs. It is visually apparent that the BPTT based design associates to the right pattern from very much distorted patterns. For the experiments seen on Figure \ref{fig:test_compare} we down-scaled the images from 28x28 to 7x7 which distorted many of the inputs but helped speeding up the computations, as an all-to-all coupled 728 oscillator system would result in almost 620000 resistors which is hard to phyiscally realize.

\begin{figure}[!ht]
    \centering
    \subfloat[Proposed nearest neighbor-connected circuit design by ML]{{\includegraphics[width=8cm]{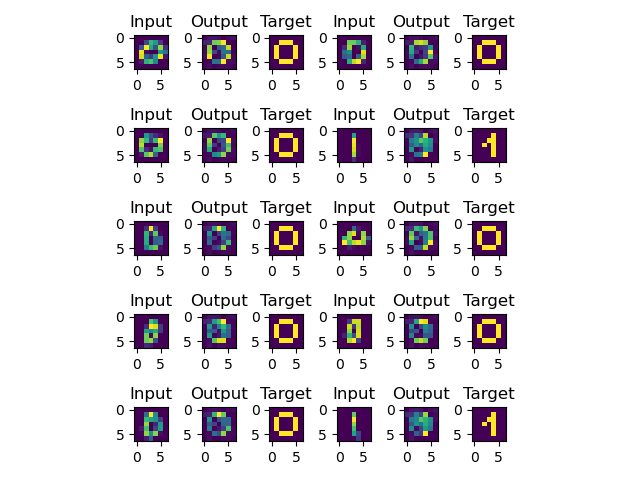} }}%
    \qquad
    \subfloat[Circuit with existing Hebbian-learning based couplings]{{\includegraphics[width=8cm]{test_hebian.png} }}%
    \caption{On a) we can see the results of the nearest neighbor connected network trained by ML, while on b) we can see the results of the system trained by the Hebbian-learning scheme. Here the qualitative results is not the clearly different, but it is at least on par with the Hebbian-based method and also in some cases it even outperforms it.}%
    \label{fig:test_compare_nn}%
\end{figure}

Most importantly, the BPTT-based design allows the design of sparsely interconnected circuit topologies. We used it to design the $C_{ij}$ matrix of associative memory assuming only nearest neighbor interconnections. The nearest-neighbor interconnected, BPTT-designed network outperforms the fully interconnected Hebbian network - even if the number of learnable parameters in the system ($\approx 8n$ vs. $\frac{1}{2}n^2$) is significantly less. The qualitative results of this comparions can be seen on Figure \ref{fig:test_compare_nn}.

Quantitatively the results of the different approaches for the whole dataset $S = \{0, 1\}$ can be seen in Table \ref{tab:qualitatives}.
\begin{table}[!h]
    \centering
    \begin{tabular}{r|ccc}
        \textbf{Method}   & \textbf{Hebbian} & \textbf{Proposed fully connected} & \textbf{Proposed NN connected} \\ \hline
\textbf{\#Params} & 1176             & 2352                              & 312                            \\
\textbf{MSE}      & 0.068            & 0.020                             & 0.047           
    \end{tabular}
    \caption{\label{tab:qualitatives} The MSEs of all the elements from the set and their respective ground truths for the different methods. It is apparent that the fully connected network performed the best, but even the nearest neighbour connected layer is good enough to beat the Hebbian learning in terms of quantitative association.}
\end{table}

\subsection{Non-associative classifiers with hidden layers}
\label{sec:classifier}
Single layer associative memories are not particularly efficient for classifying all the 10 MNIST classes, as there are strong correlations between the different digits. For this reason we also investigated multi-layered ONNs for this task. 

\begin{figure}[!ht]
    \centering
    \includegraphics[width=0.8\textwidth]{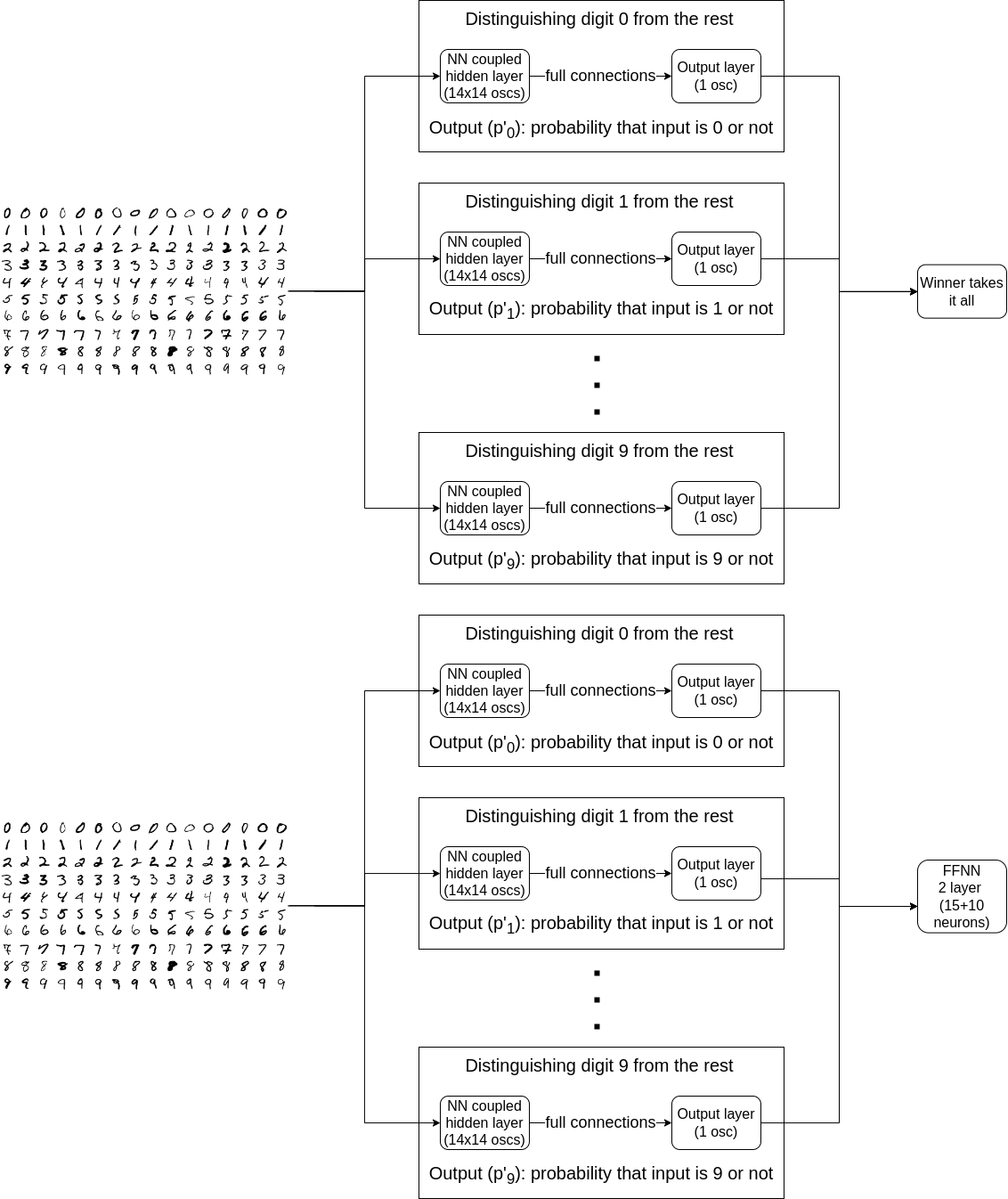}
    \caption{Here it can be seen the two of the three tested architecture for the time-independent MNIST classification. Both consists the individually trained, nearest neighbour connected subnetworks which were designed to distinguish between a single class and the rest of the classes using binary cross-entropy loss function. The top block diagram describes the algorithm where to pick the prediction we took the maximum of individual network output probabilities. The more sophisticated version can be seen on the bottom block diagram. Here we took the output probabilities of the individual classifiers and feed them as inputs to a small, regular FFNN and trained it as it were a 10-class classification problem using binary cross-entropy.}
    \label{fig:MNIST_architectures}
\end{figure}

First we started with the binary classification task which was easier to solve, so the regular, single-hidden layered, 1 output neuron setup was sufficient to solve it, but for the multi-class prediction, we created three different architecures: one with a structure resembling that of a regular 1-hidden layered, 10-output feed forward neural network (FFNN) but only from oscillators as neuron, but having a fully interconnected hidden layer; one where we had only a single output oscillator, but the same hidden layer and we created 10 of these and each individual apparatus was designed to distinguish one digit from the rest and then a winner takes it all model decides which class the input belongs to; and also, we modified the previous one so the output of the system was not the probabilities used for the winner takes it all algorithm, but rather they were fed to a small, regular neural network with 15 neurons in the hidden layer and 10 neurons in the output layer as a multiclass classification would require, see on Fig \ref{fig:MNIST_architectures}.

The first approach to the multi-class prediction with the regular, FFNN-like structure was resulted in a 70-75\% predictive accuracy, which was not satisfactory, so we omit the details here but discuss the other two approaches in later sections.

\subsubsection{Binary classifiers with a single output}

\begin{figure}[!ht]
    \centering
    \includegraphics[width=0.8\textwidth]{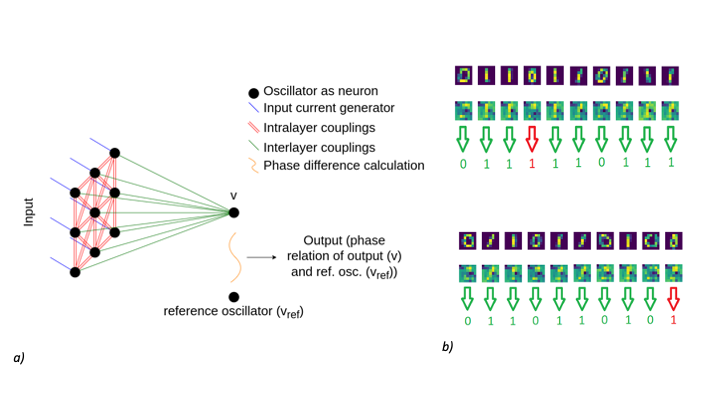}
    \caption{A simple two-layer classifier showing also the patterns forming in the hidden layer. }
    \label{fig:singleout}
\end{figure}

The two-layer classier is is shown in Fig. \ref{fig:singleout}. the phase of the output oscillator carrier the classification result: we compare the output oscillator phase with a reference oscillator and maximize (minimize) their phase difference for one (or the other) pattern.

Since the optimal oscillator couplings are discovered by the BPTT algorithm, this device  does not necessarily work as an associative memory. The phase patterns appearing in the hidden layer are non-intuitive, even if they can vaguely resemble to the images being recognized.

That having been said, without any apparent, clearly visible structure in the hidden layer, the network was predicting the two classes at a 98\% success rate. The predictions made on some images are present in Fig. \ref{fig:singleout}.

\subsubsection{10-digit classifier using a winner take all output}

\begin{figure}[!ht]
    \centering
    \includegraphics[width=\linewidth]{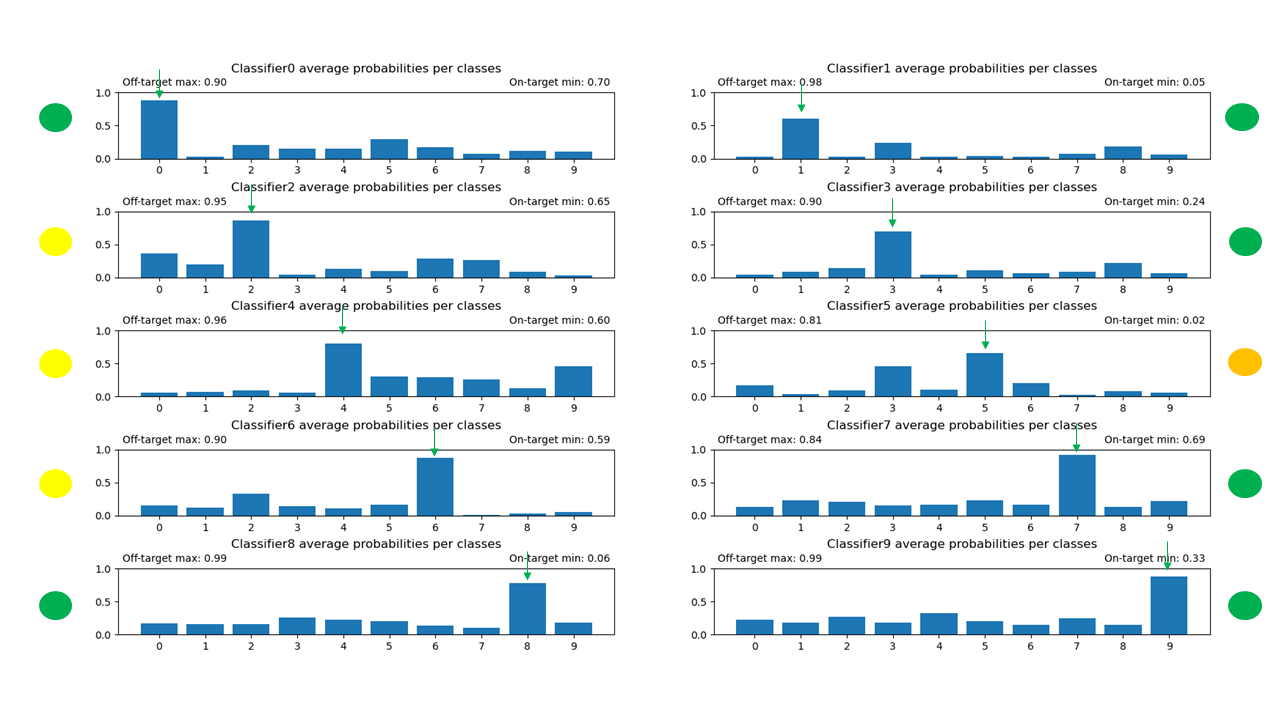}
    \caption{The distribution of predicted average probabilities for the individual, competitive networks that were used in the winner takes it all model.}
    \label{fig:winnertakesitall}
\end{figure}

Moving on to the winner takes it all architecture, the results of the distribution of average values of the predictions of each individual, competitive network can be seen on Fig \ref{fig:winnertakesitall}. Each subnetwork is resposible for recognizing one particular digit - albeit as seen in Fig \ref{fig:winnertakesitall} it often predicts high likelihood for the wrong class. Using the winner take all algorithm (i.e. the digit is identified by the ONN network giving the highest output) we achieve an accuracy around 70 \% - far better than random guessing (that would be 10 \%) but far from a good result.

%when we measured the accuracy we got 65-70\%, which was worse than the fully connected, initial idea, but it has to be noted that here we had nearest neighbour connection in the hidden oscillatory layer instead of a fully connected one, so we moved forward with this, but used the modified, more complex "post-processing" apparatus as well, namely the small FFNN. 

%\begin{figure}[!ht]
%    \centering
    %\includegraphics[width=\linewidth]{winnertakesitall.png}
    %\caption{The distribution of predicted average probabilities for the individual, competitive networks in the winner takes it all model. The red circled bars are those that on average was too high as probabilities because the given subnetwork should not have such high value for that specific digit. The green arrows indiciate which bar should be the highest. The yellow, orange and red dots near the plots indiciate how well the subnetwork managed to solve its task. it can be seen that this had to be improved.}
    %\label{fig:winnertakesitall}
%\end{figure}

\subsubsection{10-digit classifier using using a trained second layer}

Instead of the winner takes all decision, we used a simple trained perceptron layer at the end to improve classification accuracy. There are only very few multi-layered ONNs in the literature \cite{ref:karg}.

Using the outputs of the competitive networks as inputs to this small neural network we managed to reach 96.7\% predictive accuracy which was good enough to test it with a similar sized - in terms of paramater count -, regular software neural network on the same dataset and that could not reach 96.7\%, just 93-95\%. It has to be noted, that this problem is solved with fairly trivial networks with better accuracy but those are using hundreds of thousands of parameters and we only had ~20000 parameters in our training scheme. 

\begin{figure}[!ht]
    \centering
    \includegraphics[width=0.8\textwidth]{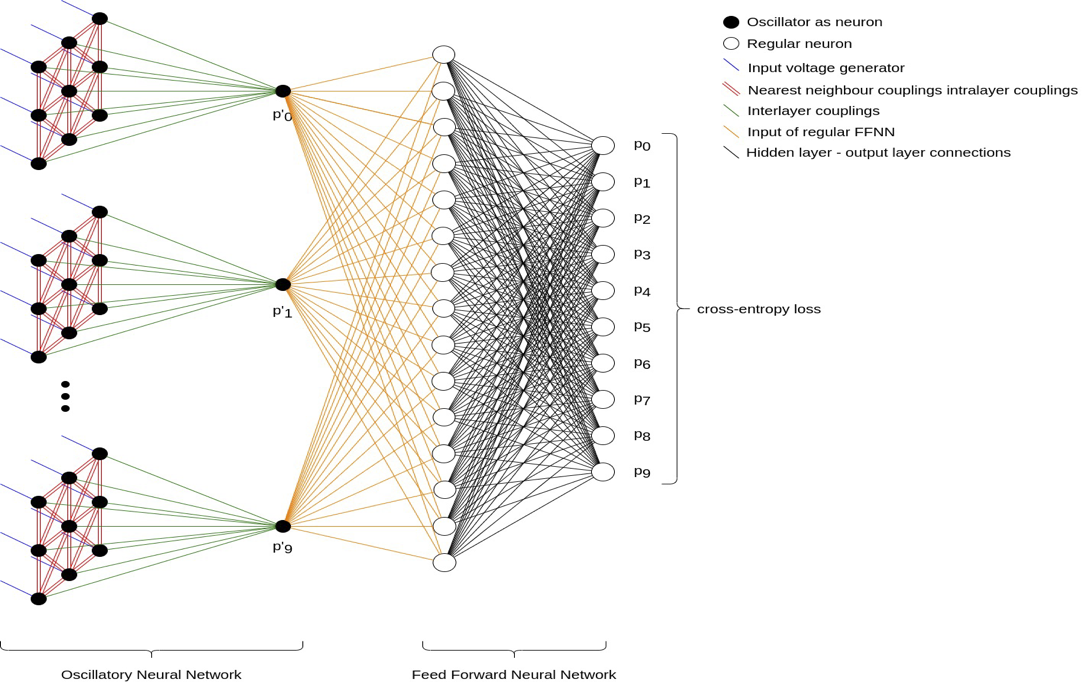}
    \caption{A network architecture with ONN layers as preprocessors and a traditional neural network postprocessing the results. The easy to train output layer significantly improves classification accuracy.}
    \label{fig:neuralout}
\end{figure}

We emphasize that in terms of computation workload, the heavy lifting in this architecture is done by the ONN preprocessing layer - the output layer contains small number of parameters and it is a very small-scale neural network by any standard. The output layer is there since it is easily trainable so it can maximize network performance at low training cost. The power consumption of the network is dominated by the ONN - and so the entire architecture benefits from the energy efficient ONN operation.

\subsection{Comparison of ONN classifier architectures}

The table below summarizes some key findings of our work. Perhaps most importantly, the ONN-based network outperforms a standard FFNN with the same parameters - and it does its job with a significantly higher power efficiency than the equaivalent digital of software implementation.

\begin{table}[!h]
    \centering
    \begin{tabular}{rcc|ccc|c}
                                         & \multicolumn{2}{c|}{\textbf{Binary classifiers}} & \multicolumn{3}{c|}{\textbf{Multiclass classifiers with oscillators}}                                & \textbf{Benchmark}                \\ \cline{2-7} 
\multicolumn{1}{r|}{\textbf{Method}}     & \textit{\textbf{Fully}}  & \textit{\textbf{NN}}  & \textit{\textbf{FFNN-like}} & \textit{\textbf{Winner takes it all}} & \textit{\textbf{FFNN 2-layer}} & \textit{\textbf{Perceptron FFNN}} \\
\multicolumn{1}{r|}{\textbf{\#Param}}    & 38416                    & 1600                  & 40180                       & 16000                                 & 16325                          & 16363                             \\
\multicolumn{1}{r|}{\textbf{Perf. (\%)}} & 98                       & 98                    & 70-75                       & 65-70                                 & 96.7                           & 93-95                            
\end{tabular}
    \caption{\label{tab:qualitatives} The quantitative comparisons of binary and multi-class classifiers with the parameter count indicated.}
\end{table}

\section*{Conclusions and outlook}

In this paper we introduced an in-silico method to design ONNs. We build a computational model of the ONN, apply BPTT techniques on this model and determine circuit parameters automatically using the BPTT training algorithm. This way we can design ONNs that are not limited by the lack of specific learning rules. The BPTT-based design allows us to explore the limits of ONN hardware without the limitations imposed by the simplicity of training algorithm.

As one of the main results of the work we find that a nearest-neighbor connected ONN that is designed by BPTT can outperform a fully connected Hebbian-trained device. This discovery opens the door to physically realizable ONNs, which perform complex processing functions without unfeasibly high number of interconnections.

Secondly, we developed multi-layered ONN devices, of which very few exist in literature. In line with expectations, we find that multiple layers significantly enhance the capabilities of the network. When the ONN first layer (preprocessing layer) is followed by a simple perceptron-based FFNN a classification accuracy of 95 \% is reached.  Most of the network complexity is in the first (preprocessing) layer, so the energy efficient operation of the ONN hardware is taken advantage of. 

\bibliography{bib}

\section*{Acknowledgements}
This work was partially supported by a grant from Intel corporation, titled HIMON: 'Hierarchically Interconnected Oscillator networks' We are grateful for regular and fruitful discussions with the Intel team in particular Narayan Srinivasa, Dmitri Nikonov and Amir Khosrowshahi.

\section*{Additional information}
\textbf{Competing financial interests:} 
The authors declare no competing financial interests.

\end{document}